\documentclass[aps,reprint,prl,longbibliography,superscriptaddress]{revtex4-1}
\usepackage{}
\usepackage{amssymb}
\usepackage{amsmath}
\usepackage{dcolumn}
\usepackage{bm}
\usepackage{graphicx}
\usepackage{mathrsfs}
\usepackage[colorlinks,linkcolor=blue,anchorcolor=blue,citecolor=blue,urlcolor=black]%
{hyperref}

\begin{document}

\title{Band-gap-engineered spin-phonon, and  spin-spin interactions  with defect centers in diamond coupled  to  phononic crystals }

\author{Peng-Bo Li}
\affiliation {Shaanxi Province Key Laboratory of Quantum Information and Quantum Optoelectronic Devices,
 Department of Applied Physics, Xi'an Jiaotong University, Xi'an
710049, China}
\author{Xiao-Xiao Li}
\affiliation {Shaanxi Province Key Laboratory of Quantum Information and Quantum Optoelectronic Devices,
 Department of Applied Physics, Xi'an Jiaotong University, Xi'an
710049, China}
\affiliation {Department of Physics, University of Oregon, Eugene, Oregon 97403, USA}

\author{Franco Nori}
\affiliation {Theoretical Quantum Physics Laboratory, RIKEN Cluster for Pioneering Research, Wako-shi, Saitama 351-0198, Japan
 }
\affiliation {Department of Physics, The University of Michigan, Ann Arbor, Michigan 48109-1040, USA}

\begin{abstract}
We study  a  spin-phononic  system where  diamond defect centers are interfaced with  a quasi-one-dimensional phononic crystal.
We show that, a single defect center, coupled to the phonon modes of a  phononic crystal waveguide near the band gap,  can seed  its own phononic cavity mode with an exponentially decaying envelope around the defect center's position. The spin-induced phononic cavity, with a greatly reduced and tunable mode volume,  allows coherent phonon-mediated interactions between distant spins with a highly tunable range, enabling access to a variety of long-range interacting spin models. This work opens  prospects for exploring quantum many-body physics and  quantum information processing with defect centers and periodic  phononic nanostructures.

\end{abstract}

\maketitle

\emph {Introduction.--} Electronic  spins associated with defect centers in
diamond comprise an outstanding platform for both fundamental research and practical
applications \cite{RepProgPhys-74-104401,NatPhoton-5-397,NatPhotonic-10-631,pt-67-38,nature-455-648,nature-455-644,nature-514-72}. Prominent defect centers in diamond include silicon-vacancy (SiV) \cite{prl-112-036405,prl-113-113602,prl-119-223602,prb-97-205444,prl-113-263602,NC-7-13512,NC-8-14451}, nitrogen-vacancy (NV) \cite{PR-528-1} and germanium vacancy (GeV) \cite{prl-118-223603}
color centers, which possess the advantage of  long coherence times \cite{Naure-Mat,NC-4-1743} and perfect compatibility with other solid-state setups \cite{SCI-339-1174,SCI-354-847,prl-105-210501,nature-478-221,prappl-4-044003,prappl-10-024011}.

For  the realization of practical quantum technologies, coherent and controllable interactions between distant
solid spins play an essential role. To achieve this goal,  schemes for interfacing solid spins via mechanical degrees of
freedom have been extensively investigated \cite{SCI-335-1603,prx-5-031031,prx-6-041060,prl-116-143602,prx-8-041027,prl-120-2136031,prl-117-015502,prl-121-123604,prl-110-156402,natphys-6-602,np-7-879,jo-19-033001,nl-12,prb-79-041302,prb-94-214115-2016,prl-111-227602,NC-9-2012}, where a  conventional acoustic
cavity (waveguide) \cite{SCI-346-207,NC-8-975,prx-8-041034,nl-12-6084} or mechanical resonator is often employed to mediate effective spin-spin interactions.  Nevertheless, there are
two challenges in complete quantum control of both spin and mechanical degrees of freedom \cite{PR-395-159,PR-511-273}.
First of all, the phonon-mediated spin-spin interaction is often too weak due to the weak intrinsic strain
coupling of spins to vibrational modes \cite{prl-113-020503,Natcomm-5-4429}. Second and more importantly, complete control on spin-spin interactions still
remains a challenge. In particular, the spatial range of  phonon-mediated spin-spin interactions cannot be tuned or controlled, and is
set solely  by the dimension of the setup.

In this work, we propose that by utilizing phononic crystals that interface with defect centers, the  problem mentioned above can be
overcome. Phononic crystals \cite{PC,prl-71-2022,nature-462-78,nature-503-209,prl-121-040501,prl-112-153603,prx-5-041051,prl-121-194301,OE-18-14926,Optica-3-1404,prappl-9-044021} naturally
enable interactions with a strength potentially much greater, owing to the
much tighter confinement of the mediating phonon. The ability to tailor the modal properties and dispersion
relation of a phononic crystal waveguide significantly goes beyond that of a  conventional acoustic
waveguide or cavity, which offers a greatly expanded toolbox for controlling
spin-phonon interactions.  Particularly, the band gap of a phononic crystal provides a tunable interaction range, a feature
which is unique to this kind of nanostructures, and makes
phononic crystals remarkably different from either acoustic cavities or
unstructured waveguides \cite{PC,prl-71-2022,nature-462-78,nature-503-209,prl-121-040501,prl-112-153603,prx-5-041051,prl-121-194301,OE-18-14926,Optica-3-1404,prappl-9-044021}. We investigate the band gap engineered spin-phonon, and spin-spin interactions with
defect centers in diamond interfaced to a quasi-one-dimensional (1D) phononic crystal. We show that, when the
transition frequency of the defect center lies within the band gap, the defect center can seed its own phononic cavity mode with an exponentially decaying envelope around its position. This leads to an enhanced spin-phonon coupling due to the
much tighter confinement of the phonon. The band-gap interaction between defect centers and phononic crystal modes  can realize coherent spin-spin coupling with a highly tunable spatial range, which is not readily achievable using other interaction mechanisms. This work indicates that hybrid systems \cite{RMP-1} composed of defect centers and periodic  phononic nanostructures comprise a promising platform for the investigation of quantum many-body physics \cite{prl-75-553} and  quantum information processing.

\emph {Model.--}
\begin{figure}[t]
\centerline{\includegraphics[bb=0 116 847 481,totalheight=1.45in,clip]{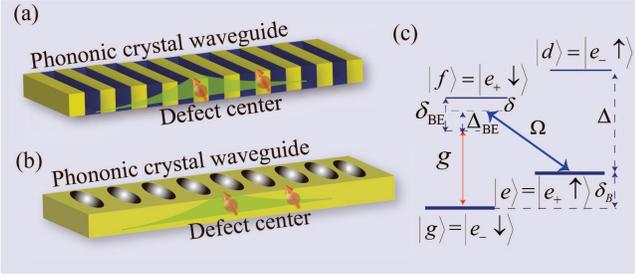}}
\caption{(Color online)  Schematic of  two separated defect centers in diamond  coupled to  the phonon modes of a 1D phononic crystal waveguide near a  bandgap.  (a) The phononic crystal waveguide is a quasi-1D periodic structure composed of alternating
layers of diamond and Si (type I). (b) The phononic crystal waveguide is a 1D phononic crystal consisting of a periodic array of holes in a diamond beam (type II). (c) Driven $\Lambda$ system of SiV centers, where the transition $\vert g\rangle\leftrightarrow \vert f\rangle$ couples with strength $g$ to the phononic crystal modes, while the transition $\vert e\rangle\leftrightarrow \vert f\rangle$ is driven by a classical microwave field with detuning $\delta$ and Rabi frequency $\Omega$.  When $\delta_B\gg\delta$, the state $\vert d\rangle$ is far off-resonance and can be neglected.
 }
\end{figure}
We consider the setup shown in Fig.~1, where separated defect centers (SiV, NV, and GeV centers) in diamond are  coupled to  the phonon modes of a 1D phononic crystal waveguide near a  band gap.  Without loss of generality, we assume that the defect center is coupled via strain to a continuum of compression modes propagating along the phononic crystal (along the $x$ axis). Local lattice
distortions associated with internal compression modes of the phononic crystal affect the defect's electronic structure, which induces a strain coupling between these phonons and the orbital degrees
of freedom of the center \cite{prb-94-214115-2016,prb-97-205444}.

The phononic crystal, a quasi-1D periodic structure, has a spatially periodic structure with lattice constant $a$, a cross section $A$,
and total length $L\gg a, \sqrt{A}$. We assume that the transverse dimensions of the phononic crystal are much smaller than the characteristic
phonon wavelength $\lambda_c= 2\pi v_c/\omega_s$, with $v_c$ the characteristic speed of sound in diamond, and $\omega_s$ the effective transition frequency of the defect center.  In this case,  the transverse  modes are far off resonance with the defect center, and
only the longitudinal waves are involved in the coupling.

The phononic crystal waveguide supports
phonon modes of frequency $\omega_{n,k}$ and mode profile $\vec{Q}_{n,k}(\vec{r})$ \footnote{See Supplemental Material for more
details}, where $k$ is the wave vector along
the waveguide, and $n$ is the band index.
Because of the periodicity of the phononic crystal, the phononic modes are of Bloch form \cite{prl-71-2022}. In the limit of a quasi-1D crystal, $\vec{Q}_{n,k}(\vec{r})\sim e^{ikx}\vec{u}_{n,k}(\vec{r})$ ($\vert k\vert\leq \pi/a$ in the first Brillouin zone). Here $\vec{u}_{n,k}(\vec{r})$ is a periodic function
associated with the shape of the Bloch modes, with periodicity given by the lattice constant $a$.
\begin{figure}[t]
\centerline{\includegraphics[bb=822 321 1195 753,totalheight=3.2in,clip]{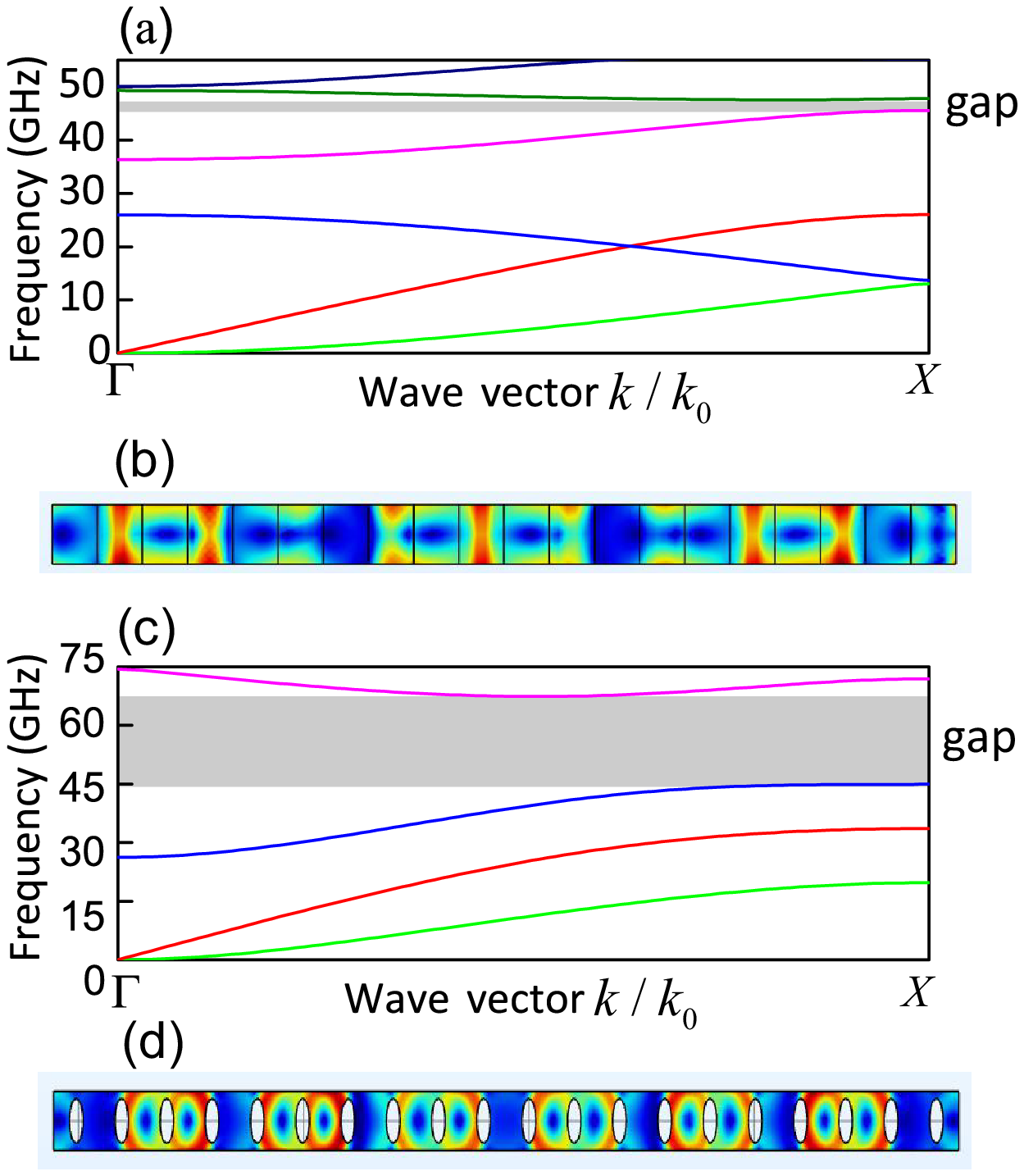}}
\caption{(Color online)  (a) Phononic band structure of the  quasi-1D phononic crystal waveguide of type I.
The cross section has a dimension $100~\text{nm}\times 20 ~\text{nm}$, and the crystal has a period of 150 nm.
(b) Displacement pattern of the symmetric  mode at the band edge frequency $\omega_\text{BE}$ for the type I phononic crystal.
(c) and (d) show the corresponding results for the type II phononic crystal, which has a width of 100 nm, and a period of 100 nm. For the elliptical holes
in the waveguide, the minor (major) axis is 30 nm (76 nm).}
\end{figure}
Figure 2 shows finite-element-method (FEM) simulations of the mechanical band structures, and displacement patterns of the symmetric modes at the band edge frequency for the two types of phononic crystal waveguides. The band structure of the phononic crystal waveguide shows a sizable band gap for the mechanical modes, which allows the defect center's transition frequency to lie within the bandgap.

For defect centers such as SiV,  NV, and GeV centers, strong coupling
between the orbital degrees of freedom and the mechanical vibrations can
be obtained through the ground states or excited states \cite{prb-94-214115-2016,prb-97-205444,prx-6-041060,prappl-6-034005}. Here we take SiV  centers
as an example, but the general model also applies to NV and GeV centers. The SiV center is a point defect in which
a silicon atom is positioned  between two adjacent
missing carbon atoms in the diamond lattice \cite{prb-97-205444}. Its electronic ground state is formed by an unpaired hole of spin $S=1/2$, which occupies one of the two degenerate orbital states
$\vert e_x\rangle$ and $\vert e_y\rangle$ \cite{Note1}. In the presence of spin-orbit interactions and
a weak Jahn-Teller effect, the four states are split into two
doublets \cite{prb-97-205444}, $\{\vert g\rangle=\vert e_-\downarrow\rangle, \vert e\rangle=\vert e_+\uparrow\rangle\}$
and $\{\vert f\rangle=\vert e_+\downarrow\rangle, \vert d\rangle=\vert e_-\uparrow\rangle\}$.
Here, $\vert e_{\pm}\rangle=(\vert e_x\rangle\pm i\vert e_y\rangle)/\sqrt{2}$ are eigenstates of
the orbital angular momentum operator $\hat {L}_z$, with the $z$ axis  along the symmetry axis of
the defect. The energy gap between these doublets is $\Delta/2\pi=46$ GHz \cite{prb-97-205444}.

The interaction Hamiltonian
is given by $H_\text{strain}=\sum_{ij}V_{ij}\epsilon_{ij}$ \cite{prb-97-205444,jo-19-033001,Note1}, where $V_{ij}$ is an operator acting on the electronic states of the defect center, and
$\epsilon$ is the strain tensor. The strain
Hamiltonian can be projected onto the irreducible representations of the D$_{3d}$ group for SiV centers, which reflects the symmetry of the orbital wavefunctions, i.e., $H_\text{strain}=\hbar\sum_{l}V_{l}\epsilon_{l}$. Each $\epsilon_{l}$ is a linear
combination of strain components $\epsilon_{i,j}$, and corresponds to
specific symmetries indicated by the subscript $l$, i.e., $\epsilon_{A_{1g}} = t_{\perp} (\epsilon_{xx}+\epsilon_{yy})+t_{\parallel}\epsilon_{zz}$,
$ \epsilon_{E_{gx}} = d(\epsilon_{xx}-\epsilon_{yy}))+f\epsilon_{zx}$, and $\epsilon_{E_{gy}} = -2d\epsilon_{xy}+f\epsilon_{zx}$.
Here $t_{\perp},t_{\parallel},d$, and $f$ are the four strain-susceptibility parameters
that completely describe the strain-response of the orbital states $\vert e_x\rangle$ and $\vert e_y\rangle$.
The strain Hamiltonian can be rewritten in the basis spanned by the
eigenstates of the spin-orbit coupling \cite{prl-120-2136031}
$H_\text{strain}=\hbar\epsilon_{E_{gx}}(L_-+L_+)-i \hbar\epsilon_{E_{gy}}(L_--L_+)$,
where $L_+=L_-^\dag=\vert f\rangle\langle g\vert+\vert e\rangle\langle d\vert$ is the orbital raising operator within the ground state.

Upon quantization, the mechanical displacement $\vec{Q}(\vec{r})$
becomes an operator, which  can be expressed
in terms of the elementary normal modes and annihilation operators as $\hat{Q}(\vec{r})=\sum_{n,k}[\vec{Q}_{n,k}(\vec{r})\hat{a}_{n,k}+\text{H.c.}]$ \cite{Note1,CavityOptM}.
The resulting strain coupling can be written in the general form $\hat{H}_\text{strain}\simeq \sum_{n,k}[\hbar g_{n,k}\hat{a}_{n,k}\hat{J}_+e^{ikx}+\text{H.c.}]$ \cite{Note1},
where the coupling constant $g_{n,k}$ depends on the local strain tensor and can be evaluated for a known mode profile $\vec{Q}_{n,k}(\vec{r})$,
and $\hat{J}_+=\vert f\rangle\langle g\vert+\vert d\rangle\langle e\vert$.
The defect center is coupled predominantly to
only a single band of the  phononic crystal waveguide, via tuning the transition frequency $\Delta$ of the color center close
to the band edge frequency $\omega_\text{BE}$, with detuning $\delta_\text{BE}=\Delta-\omega_\text{BE}$ (where $\delta_\text{BE}>0,$ so that the spin frequency lies within the band
gap). We assume that the detuning to any other band edge is much
larger than $\delta_\text{BE}$, so the other bands can be neglected. For clarity, we omit the band index $n$ in the following discussion.

\emph {Phononic bound state and effective  acoustic cavity.--}
When the transition frequency of the
defect center is tuned close to the band edge, the spin is dominantly coupled to the modes
near the band edge wavevector $k_0=\pi/a$ due to the van Hove singularity in the density of states, i.e., $\frac{\partial k}{\partial \omega_k}|_{k_0}\rightarrow \infty$.
In this case, the dispersion relation can be approximated to be quadratic $\omega_k\simeq \omega_\text{BE}-\alpha a^2(k-k_0)^2$, with
$\alpha$ a parameter characterizing the band curvature.  In the presence of a static magnetic field
$\vec{B}=B_0\vec{e}_z$ and a microwave driving field of frequency $\omega_0$ and Rabi-frequency $\Omega$ along $\vec{e}_x$, we can
implement a Raman transition between the states $\vert g\rangle$ and $\vert e\rangle$ via the excited state $\vert f\rangle$ through coupling the SiV center to the mechanical modes.
The Hamiltonian of the whole system can be written as \cite{Note1}
\begin{eqnarray}
  \hat{\mathcal {H}} &=& \sum_{k} \hbar \omega_k \hat{a}^\dag_k\hat{a}_k+\hbar \omega_s \hat{\sigma}_{ee}\\ \nonumber
  &&+\sum_{k} \hbar g_\text{eff}( \hat{a}^\dag_k
  \hat{\sigma}_{ge}e^{-ikx_0}+\hat{a}_k \hat{\sigma}_{eg}e^{ikx_0}),
\end{eqnarray}
where $\hat{\sigma}_{ij}=\vert i\rangle\langle j\vert$, $\omega_s=\Delta-\delta$, $g_\text{eff}=g_k\Omega/\delta$, and
$g_k$ is the coupling strength between the defect center (with the position $\vec{r}_0$) and phonon modes in the band. Moreover, for the modes near the band edge the coupling strength $g_k$ is approximately independent of $k$, i.e., $g_k\sim  g\sim \frac{d}{v_l}\sqrt{\hbar \omega_\text{BE}/2\pi\rho aA}$ \cite{Note1}, where $d/2\pi\sim 1$ PHz/strain is the strain sensitivity, and $v_l=1.71\times10^4~\text{m/s}$  is the speed of sound in diamond.

For a single excitation in the system,
there exists a bound state $\vert \varphi_b\rangle=\cos \theta\vert 0\rangle\vert e\rangle+\sin\theta\vert 1\rangle\vert g\rangle$
within the bandgap with the eigenenergy $\hbar\Omega_b$. Here $\vert 0\rangle$ is the vacuum state for the phonon modes, and
$\vert 1\rangle=\int dk c_k \hat{a}_k^\dag\vert0\rangle$ is a single phonon excitation of the modes in the band. The bound state and
the corresponding eigenenergy are determined by the eigenvalue equation $\hat{\mathcal {H}}\vert \varphi_b\rangle=\hbar\Omega_b\vert \varphi_b\rangle$. Solving this equation yields \cite{Note1}
\begin{eqnarray}
 \Omega_b-\omega_s &=&\frac{\pi g_\text{eff}^2}{\sqrt{\alpha (\Omega_b-\omega_\text{BE})}} \nonumber\\
 \tan^2(\theta)&=&\frac{\omega_s-\Omega_b}{2(\omega_\text{BE}-\Omega_b)}\nonumber\\
 c_k&=& \frac{g_\text{eff}e^{-ikx_0}}{\tan\theta(\Omega_b-\omega_k)}.
\end{eqnarray}
In real space, the phononic part of the bound state is exponentially localized around the spin
with spatial mode envelope
\begin{eqnarray}
\mathcal {E}(x)=\int dk c_k Q_k(x)=\sqrt{\frac{2\pi}{L_{c}}}e^{-\vert x-x_0\vert/L_{c}}Q_{k_0},
\end{eqnarray}
where the localization length $L_c$ is given by
\begin{eqnarray}
L_c&=&a\sqrt{\frac{\alpha}{\Omega_b-\omega_\text{BE}}}.
\end{eqnarray}
This localized phononic cloud has the same properties as a confined acoustic cavity mode.
The localization of the phononic wavefunction is determined predominantly by the properties of the band edge, and the frequency detunings, leading to the possibility of dynamic tuning of the localization length.

The coupling between the defect center and the phononic crystal modes near the band edge can be well understood by mapping to the
Jaynes-Cummings model  in cavity quantum-electrodynamics \cite{prl-64-2418,RMP-2} or spin-mechanics.
In particular, the eigenstate $\vert \varphi_b\rangle$ looks identical in form to one of the dressed
eigenstates  of the Jaynes-Cummings model. This mapping can be made formal: the phonon confined
around the defect center has the same functionality as the mode of
an actual acoustic cavity.
We can associate an effective spin-phonon coupling strength
$g_c=g_\text{eff}\sqrt{2\pi a/L_c}$ with the bandgap setup.
The effective vacuum-Rabi splitting $g_c$ is
identical to that of a real cavity whose mode volume is the same
as the bound phonon size $V_\text{eff}=AL_c$. This effective spin-phonon interaction can be
much stronger than that of defect centers coupling to a conventional phononic waveguide \cite{prl-120-2136031},
via tuning the effective acoustic cavity mode volume to a much smaller value such that the phonons are confined much tighter.
The effective acoustic cavity frequency is the average frequency of the phonon modes, i.e., $\int dk \vert c_k\vert^2\omega_k=\omega_\text{BE}-\delta_b$, and
the effective spin-cavity detuning is $\Delta_\text{E}=\Delta_\text{BE}+\delta_b$.

\begin{figure}[t]
\centerline{\includegraphics[bb=822 426 1104 650,totalheight=2.7in,clip]{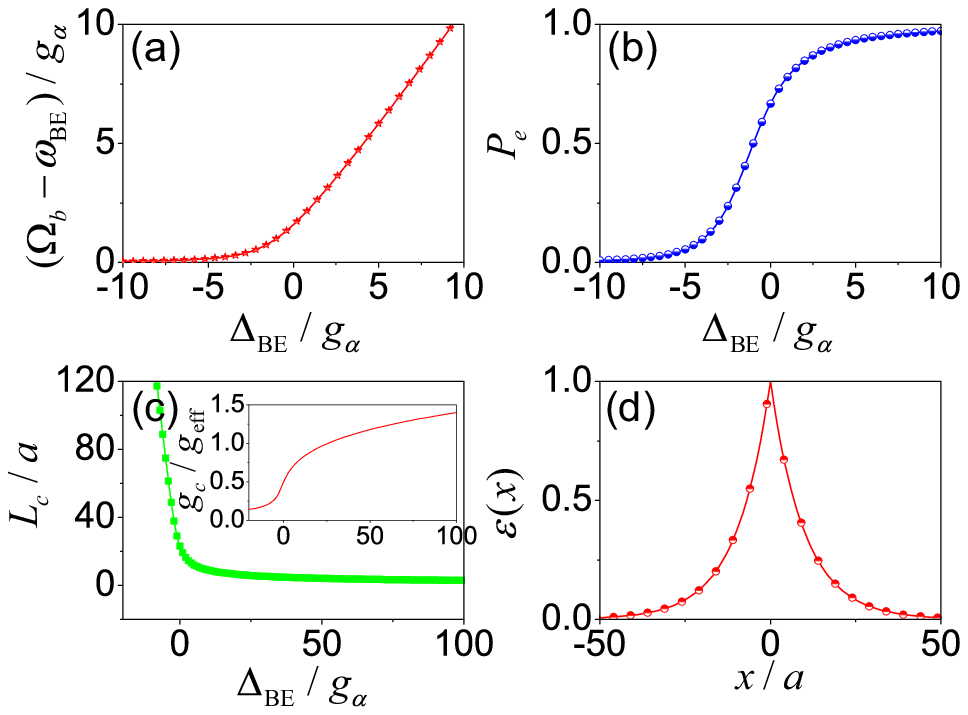}}
\caption{(Color online)  (a) The bound-state eigenfrequency  $\Omega_b$ relative to the band edge  versus detuning $\Delta_\text{BE}=\delta_\text{BE}-\delta$.
(b) The defect center's excited state population  $P_e=\cos^2\theta$ as a function of detuning $\Delta_\text{BE}$.
 (c) The length of the effective phononic cavity $L_c$ versus detuning $\Delta_\text{BE}$. The inset shows the coupling strength versus detuning $\Delta_\text{BE}$.
 (d)  Spatial profile of the phononic wave function $\mathcal {E}(x)$, for which we choose $L_c=10a$ and $x_0=0$. In the simulations, we choose the parameters for the type I phononic crystal as
 $\omega_\text{BE}/2\pi=45.5$ GHz, $\alpha/2\pi=3.5$ GHz, $a=150$ nm, $A=100~\text{nm}\times20~\text{nm}$, $g_\text{eff}=0.1g$, and $g_{\alpha}=(\pi g_\text{eff}^2/\sqrt{4\alpha})^{2/3}$.
 }
\end{figure}
In Fig.~3(a)-(c), we present the numerical simulations for the dependence of the bound-state eigenfrequency  $\Omega_b$,  the defect center's excited state population $P_e=\cos^2\theta$, and the length of the effective phononic cavity $L_c$ on  detuning $\Delta_\text{BE}$.
Fig.~3(a) shows that when $\Delta_\text{BE}/g_{\alpha}\ll-1$, then $\Omega_{b}$ approaches $\omega_\text{BE}$, while $\Omega_{b}$ approaches
$\omega_\text{BE}+\Delta_\text{BE}$ when $\Delta_\text{BE}/g_{\alpha}\gg 1$. Fig.~3(b) shows that when $\Delta_\text{BE}/g_{\alpha}\ll-1$,
the state becomes mostly phononic: a cavity mode dressed by the spin.
From Fig.~3(c), one can see that localized phonon arises with larger
detuning $\Delta_\text{BE}$ or flatter bands, giving rise to an enhancement of the effective spin-phonon coupling.
From Fig.~3(d), we find that  the phononic component of the hybrid spin-phonon state extends over multiple
sites.

\emph {Spin-spin interactions mediated by phononic crystals.--}
The band-gap interaction allows us to use the phononic crystal
modes as a quantum bus to perform more complex tasks. Analogous
to the spin-exchange interaction in cavity QED, two  defect centers that are separated by a distance
on the order of the length $L_c$ can exchange a phonon excitation via the induced acoustic cavity mode.
Based on this effective interaction, it is possible to implement quantum information
processing and a variety of  interacting spin models \cite{prl-75-553} with distant defect spins.

We consider two separated SiV centers coupled to the same phononic crystal modes near the band edge, through
Raman transitions between the states $\vert g\rangle$ and $\vert e\rangle$. In the interaction picture, we can obtain
the interaction between the defect centers and phononic crystal modes
$\hat{H}= \sum_{j=1,2}\sum_{k}\hbar g_\text{eff}^j( \hat{a}^\dag_k
  \hat{\sigma}_{ge}^je^{-i\delta_k t-ikx_j}+\hat{a}_k \hat{\sigma}_{eg}^je^{i\delta_k t+ikx_j})$,
with $\delta_k=\omega_s-\omega_k$. In the dispersive regime $\delta_k\gg g_\text{eff}^j$, this will lead to an effective  spin-spin interaction via the exchange of virtual phonons \cite{Note1},
\begin{eqnarray}
 \hat{\mathcal {H}}_{s-s}&=& \hbar J_{12}\hat{\sigma}_{eg}^1\hat{\sigma}_{ge}^2+\text{H.c.}\\ \nonumber
&=& \hbar \lambda_\text{eff}e^{-\vert x_1-x_2\vert/L_c}\hat{\sigma}_{eg}^1\hat{\sigma}_{ge}^2+\text{H.c.},
\end{eqnarray}
with the effective coupling strength $\lambda_\text{eff}= g_c^2/2\Delta_\text{BE}$.
Here the spatial range reflects the shape of the localized phononic cloud.

Different from the dipole-dipole interactions mediated by a conventional cavity, the form of the band-gap-mediated interactions
between spins is to decay exponentially with spin separation \cite{pra-87-033831}, as shown in Fig.~4.
Furthermore, the range of the interaction can be tuned through the effective
interaction length $L_c$, the length scale of which  can be
on the order of the length of experimental setups, and thus effectively
long-range over the system size. Therefore, the band gap interaction can realize coherent spin-spin interactions
with a highly tunable range, which cannot be readily achieved
using other interaction mechanisms.
This potentially enables the study of a variety
of interacting spin models \cite{Note1}, and  the deterministic generation of
entanglement and quantum state transfer between spins at a distance of several lattice constants though a phononic channel.
\begin{figure}[t]
\centerline{\includegraphics[bb=840  444 1062 613,totalheight=1.6in,clip]{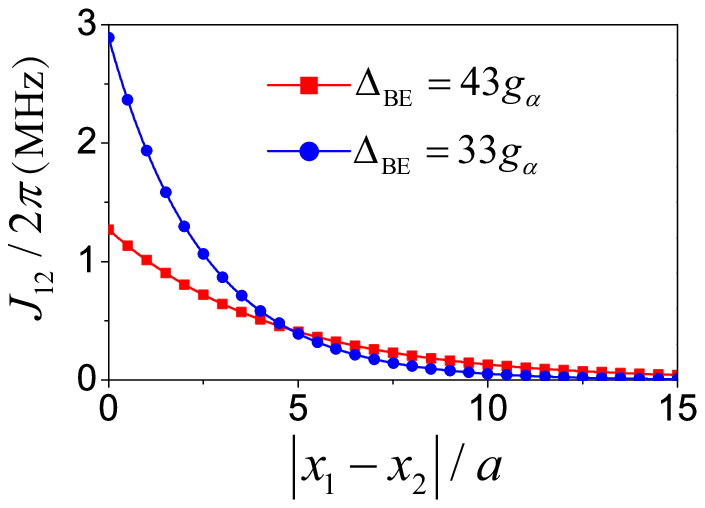}}
\caption{(Color online)  The coupling strength of  band gap mediated dipole-dipole interactions $J_{12}$
 versus $\vert x_1-x_2\vert$ for different detunings $\Delta_\text{BE}$. Other parameters are chosen as those in Fig.~3.}
\end{figure}

We assume that the two defect centers are initially prepared in the state $\psi(0)=\vert g\rangle_1\vert e\rangle_2$.
Then under the Hamiltonian $\hat{\mathcal {H}}_{s-s}$, the state evolution of the system is given by
$  \psi(t) = \cos(J_{12}t) \vert g\rangle_1\vert e\rangle_2-i\sin(J_{12}t)\vert e\rangle_1\vert g\rangle_2,$
which is an entangled state for the two centers.
If we choose $ J_{12}\tau=\pi/4$, we can obtain the maximally entangled two-particle
state $ \psi(\tau)  = \frac{1}{\sqrt{2}}(\vert g\rangle_1\vert e\rangle_2-i\vert e\rangle_1\vert g\rangle_2)$,
which is the well-known Einstein-Podolsky-Rosen state.
The interaction (5) between
the two defect centers can also be used to transfer arbitrary quantum
information encoded in ground spin states from one center to the other:
$(\alpha\vert g\rangle_1+\beta\vert e\rangle_1)\vert e\rangle_2\rightarrow (\alpha\vert g\rangle_2+\beta\vert e\rangle_2)\vert e\rangle_1$.

\begin{figure}[t]
\centerline{\includegraphics[bb=795  419 1075 651,totalheight=2.8in,clip]{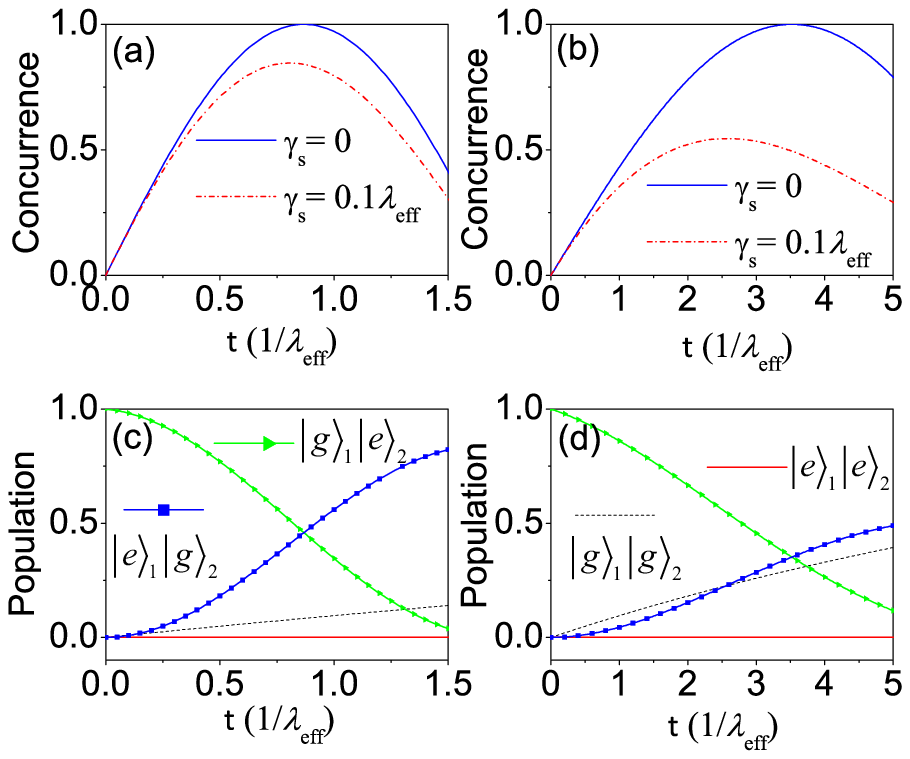}}
\caption{(Color online)  Time evolution of the concurrence and populations. For (a) and (c) the distance between the
two defect centers is $\vert x_1-x_2\vert=0.1L_c$, while for (b) and (d) the distance between them
 is $\vert x_1-x_2\vert=1.5L_c$. The results given in (c) and (d) are for
$\gamma_s=0.1\lambda_\text{eff}$.}
\end{figure}
Fig.~5 displays the numerical results for the time evolution of the concurrence and populations for two defect centers
under ideal and realistic conditions through solving the master equation using the
QuTiP library \cite{CPC}. We find that, the concurrence and populations are significantly affected by the distance between the centers, even though the other conditions are the same.  Moreover, if the decoherence of the defect centers is taken into account, for
distances between the centers larger than $L_c$, the produced entangled states and the quantum state transfer efficiency will be significantly spoiled.

For the setups illustrated in Fig.~1, the band edge frequency is $\omega_\text{BE}/2\pi=45.5$ GHz,
and the quality factor of a phononic crystal mode can reach $Q_m\sim 10^7$, resulting in a mechanical damping rate $\gamma_m/2\pi\sim 5$ kHz. This high mechanical quality factor should be achievable with the use of
phononic crystal shields and with the state-of-the-art nanofabrication technologies \cite{prx-5-041051,prx-8-031007,proceedings-1}.
At mK temperatures, the thermal phonon number is far below 1 and the spin dephasing rate of  SiV centers is about $\gamma_s/2\pi \sim 0.1$ kHz with dynamical decoupling \cite{prl-119-223602}. Even taking into account the surface effect on  SiV centers' decoherence, the spin dephasing rate is still expected to be about $\gamma_s/2\pi \sim100$ kHz.
The coupling strength for the design in Fig.~1(a) can be calculated as $g/2\pi\sim 178$ MHz. If we choose $\Omega\sim g$, $\delta\sim 10 g$, and
$\Delta_\text{BE}=43g_{\alpha}$, then we have $g_\text{eff}\sim 0.1 g$, $g_{c}/2\pi\sim 21.2$ MHz, and $\lambda_\text{eff}/2\pi\sim 1.27$ MHz.
The time for generating the entangled state and implementing quantum state transfer is about $\tau\leq 1\mu$s, which is  smaller than the coherence time of the setup.

\emph {Conclusions.--} We have investigated the band gap engineered spin-phonon, and spin-spin interactions with defect centers in
diamond coupled to a quasi-1D phononic crystal. We show that, when the transition frequency of the defect center lies
within a band gap, the defect center can seed its own phononic cavity mode with an exponentially
decaying envelope around its position. The band-gap interaction allows coherent phonon-mediated interactions between spins with a tunable spatial range. The scheme presented in this work is general and can be applied to other defect centers or solid-state systems such as SiC-based systems \cite{nature-479-84} and hexagonal boron nitride emitters \cite{NatNanotechnol-11-37}.  This work opens  prospects for exploring quantum many-body physics and quantum information processing with defect centers and  phononic
nanostructures.

X.X.L. acknowledges helpful discussions with Mark C. Kuzyk and Prof. Hailin Wang. P.B.L is supported by the NSFC under
Grant Nos. 11774285, 91536115 and 11534008, and the
Fundamental Research Funds for the Central Universities.  F.N. is supported in part by the:
MURI Center for Dynamic Magneto-Optics via the
Air Force Office of Scientific Research (AFOSR) (FA9550-14-1-0040),
Army Research Office (ARO) (Grant No. Grant No. W911NF-18-1-0358),
Asian Office of Aerospace Research and Development (AOARD) (Grant No. FA2386-18-1-4045),
Japan Science and Technology Agency (JST) (via the Q-LEAP program, the ImPACT program, and the CREST Grant No. JPMJCR1676),
Japan Society for the Promotion of Science (JSPS) (JSPS-RFBR Grant No. 17-52-50023, and JSPS-FWO Grant No. VS.059.18N),
the RIKEN-AIST Challenge Research Fund, and the
John Templeton Foundation.

%

\onecolumngrid

\appendix

\clearpage

\section*{Supplemental Material:  }


\setcounter{equation}{0}
\setcounter{figure}{0}
\setcounter{table}{0}
\setcounter{page}{1}
\makeatletter
\renewcommand{\theequation}{S\arabic{equation}}
\renewcommand{\thefigure}{S\arabic{figure}}
\renewcommand{\bibnumfmt}[1]{[S#1]}
\renewcommand{\citenumfont}[1]{S#1}

\subsection{Phononic crystals}
\subsubsection{Calculation of mechanical compression modes in diamond phononic crystals}
In the main text, we study  a quasi-1D periodic structure  with lattice constant $a$, a cross section $A$,
and total length $L\gg a, \sqrt{A}$. With the framework of elastic mechanical theory,  mechanical modes can be treated as a
continuum field with time-dependent displacement at the position $\vec{r}$, given by $\vec{Q}(\vec{r},t)$. The field
$\vec{Q}(\vec{r},t)$ obeys the equation of motion \cite{book-1}
\begin{eqnarray}
  \rho \frac{\partial^2}{\partial t^2} \vec{Q}(\vec{r},t)&=& (\lambda+\mu)\nabla(\nabla\cdot\vec{Q}(\vec{r},t))+\mu\nabla^2\vec{Q}(\vec{r},t).
\end{eqnarray}
Here $\rho$ is the mass density,  and $\lambda$ and $\mu$ are the Lam\'{e} constants.
Waves in periodic structures are best understood in terms of band theory. From Bloch's
theorem,  we have that in a structure with periodic variation, the time-harmonic solutions
to the mechanical wave equations can be expressed as a product of a
plane-wave solution $e^{ikx}$ and a periodic function $\vec{u}_{n,k}(\vec{r})$, i.e., $\vec{Q}_{n,k}(\vec{r}) = e^{ikx}\vec{u}_{n,k}(\vec{r})$,
with the eigenfrequencies $\omega_{n,k}$. The Lam\'{e} constants can be expressed in terms of the Young's modulus $E$ and
the Poisson ratio $\nu$
\begin{eqnarray}
 \lambda =\frac{\nu E}{(1+\nu)(1-2\nu)}, \qquad\mu =\frac{E}{2(1+\nu)}.
\end{eqnarray}
The frequencies $\omega_{n,k}$ and field patterns of the normal modes can be
determined by solving the corresponding eigenvalue equations using finite element
simulations \cite{Comsol}. The material properties of diamond we used are $E=1050$ GPa,
$\nu=0.2$, and $\rho=3539 ~\text{kg}/\text{m}^3$, while for silicon the material properties are
$E=170$ GPa, $\nu=0.28$, and $\rho=2329 ~\text{kg}/\text{m}^3$.
All modes considered in this work have even mirror symmetry under reflection $R_z$ about the plane
perpendicular to the axis $z$.

\subsubsection{Quantization of the displacement fields in diamond phononic crystals }
The motion can be quantized following an approach similar to that used for the electromagnetic
field in quantum optics \cite{CavityOpt}. We define phonon creation and annihilation operators $\hat{a}_{n,k}^\dag$
and $\hat{a}_{n,k}$ respectively, for each modal solution $\vec{Q}_{n,k}(\vec{r})$ of the equations of elasticity in the
structure. Then the field operator can be expressed as
\begin{eqnarray}
\hat{Q}(\vec{r})&=&\sum_{n,k}[\vec{Q}_{n,k}(\vec{r})\hat{a}_{n,k}+\text{H.c.}].
\end{eqnarray}
To calculate the proper normalization of the field profiles, we assume a single-phonon Fock state
$\vert \Psi\rangle=\hat{a}_{n,k}^\dag\vert 0\rangle$ and calculate the expectation value of additional field energy
above vacuum. We find
\begin{eqnarray}
  E_\text{mech} &=& 2\omega_{n,k}^2\int d\vec{r}\rho(\vec{r})\vec{Q}_{n,k}^*(\vec{r})\cdot\vec{Q}_{n,k}(\vec{r})\nonumber\\
  &=&2\rho V_\text{eff}\omega_{n,k}^2 \text{max}[\vert \vec{Q}_{n,k}\vert^2],
\end{eqnarray}
where the last equality defines the effective mode volume
for mode $\{n,k\}$. Setting $Q_0=\text{max}[\vert \vec{Q}_{n,k}\vert] $ and assuming the
phonon energy as $E_\text{mech}=\hbar \omega_{n,k}$, we arrive at the general
result for a phonon mode, $Q_0=\sqrt{\hbar/2\rho V_\text{eff}\omega_{n,k}}$.

\subsection{SiV centers}
As discussed in the main text, the SiV center is a point defect in which
a silicon atom is positioned  between two adjacent
missing carbon atoms in the diamond lattice. Its electronic ground state is formed by an unpaired hole of spin $S=1/2$, which occupies one of the two degenerate orbital states $\vert e_x\rangle$ and $\vert e_y\rangle$. The spin and
orbital degeneracy is lifted by the spin-orbit coupling and by the Jahn-Teller  effect.
In the presence of an external magnetic field $\vec{B}$, the Hamiltonian for the electronic
ground state of the SiV center is \cite{prb-94}
\begin{eqnarray}\label{Q1}
  \hat{H}_\text{SiV} &=& -\hbar  \lambda_\text{SO}\hat{L}_z\hat{S}_z+\hat{H}_\text{JT}+\hbar\Upsilon_LB_z\hat{L}_z+\hbar\Upsilon_s\vec{B}\cdot\vec{S},
\end{eqnarray}
where $\hat{L}_z$ and $\hat{S}_z$ are the projections of the  angular
momentum and spin operators onto the symmetry
axis of the defect center (along $z$ axis). $\lambda_\text{SO}$ is the spin-orbit coupling strength  while $\Upsilon_L$ and $\Upsilon_s$
are the orbital and spin gyromagnetic ratios respectively.  In the basis defined  by the degenerate eigenstates
$\{\vert e_x \uparrow\rangle, \vert e_x \downarrow\rangle,\vert e_y \uparrow\rangle,\vert e_y \downarrow\rangle\}$,  the different contributions
to the SiV energy levels, introduced in Eq. (\ref{Q1}), read
\begin{eqnarray}
  (\Upsilon_sB_0-\lambda_\text{SO}\hat{L}_z)\hat{S}_z &=& \frac{1}{2}\left[
                                                                             \begin{array}{cc}
                                                                              \Upsilon_sB_0 & i\lambda_\text{SO} \\
                                                                               -  i\lambda_\text{SO}& \Upsilon_sB_0 \\
                                                                             \end{array}
  \right]\otimes\left[
                  \begin{array}{cc}
                    1 & 0 \\
                    0 & -1 \\
                  \end{array}
                \right]\\
\hat{H}_\text{JT}&=&\hbar\left[
                      \begin{array}{cc}
                        \mathcal {K}_x & \mathcal {K}_y \\
                        \mathcal {K}_y & -\mathcal {K}_x \\
                      \end{array}
                    \right]\otimes\left[
                  \begin{array}{cc}
                    1 & 0 \\
                    0 & 1 \\
                  \end{array}
                \right].
\end{eqnarray}
In the above equation $\mathcal {K}_i$  represents an energy shift due to local strain in
the crystal along axis $i$. Diagonalizing Eq. (\ref{Q1}) leads to the
eigenstates
\begin{eqnarray}
  \vert g\rangle &=& (\cos\theta\vert e_x\rangle-ie^{-i\phi}\sin\theta\vert e_y\rangle )\vert \downarrow\rangle\\
   \vert e\rangle &=& (\cos\theta\vert e_x\rangle+ie^{i\phi}\sin\theta\vert e_y\rangle )\vert \uparrow\rangle\\
   \vert f\rangle &=& (\sin\theta\vert e_x\rangle+ie^{-i\phi}\cos\theta\vert e_y\rangle)\vert \downarrow\rangle \\
   \vert d\rangle&=& (\sin\theta\vert e_x\rangle-ie^{i\phi}\cos\theta\vert e_y\rangle)\vert \uparrow\rangle,
\end{eqnarray}
with
\begin{eqnarray}
    \tan\theta &=& \frac{2\mathcal {K}_x+\Delta}{\sqrt{\lambda^2_\text{SO}+4\mathcal {K}_y^2}}, \quad \tan\phi=\frac{2\mathcal {K}_y}{\lambda_\text{SO}}.
\end{eqnarray}
The corresponding eigenenergies are
\begin{eqnarray}
E_{f,g}=(-\Upsilon_sB_0\pm\Delta)/2, \quad E_{d,e}=(\Upsilon_sB_0\pm\Delta)/2
\end{eqnarray}
with $\Delta\simeq2\pi\times 46$ GHz. In this work,  we can neglect the small distortions of the orbital states
by the JT effect and in the
remainder of this work, use the approximation $\vert g\rangle\simeq\vert e_{-}\downarrow\rangle,\vert e\rangle\simeq\vert e_{+}\uparrow\rangle,\vert f\rangle\simeq\vert e_{+}\downarrow\rangle,\vert d\rangle\simeq\vert e_{-}\uparrow\rangle$.

\subsection{Strain coupling of SiV centers  to phononic crystal modes}

As discussed in the main text,  local lattice distortions associated with internal compression modes of the phononic crystal affect
the defect's electronic structure, which induces a strain coupling between these phonons and the orbital degrees of freedom of the center.
The interaction Hamiltonian
is given by
\begin{eqnarray}
H_\text{strain}=\sum_{ij}V_{ij}\epsilon_{ij}
\end{eqnarray}
where $V_{ij}$ is an operator acting on the electronic states of the defect center, and
$\epsilon$ is the strain tensor defined by
\begin{eqnarray}
\epsilon_{ij}=\frac{1}{2}\left(\frac{\partial Q_i}{\partial x_j}+\frac{\partial Q_j}{\partial x_i}  \right).
\end{eqnarray}

In this work we assume that the symmetry axis of the defect center is along the $z$ direction while the phononic crystal
is along the $x$ axis.  The strain
Hamiltonian can be projected onto the irreducible representations of the D$_{3d}$ group for SiV centers, which reflects the symmetry of the orbital wavefunctions, i.e.,
\begin{eqnarray}
  H_\text{strain}=\hbar\sum_{l}V_{l}\epsilon_{l}.
\end{eqnarray}
Each $\epsilon_{l}$ is a linear
combination of strain components $\epsilon_{i,j}$, and corresponds to
specific symmetries indicated by the subscript $l$, i.e.,
\begin{eqnarray}
  \epsilon_{A_{1g}} &=& t_{\perp} (\epsilon_{xx}+\epsilon_{yy})+t_{\parallel}\epsilon_{zz} \nonumber\\
  \epsilon_{E_{gx}} &=& d(\epsilon_{xx}-\epsilon_{yy}))+f\epsilon_{zx} \nonumber\\
  \epsilon_{E_{gy}} &=& -2d\epsilon_{xy}+f\epsilon_{yz}.
\end{eqnarray}
Here $t_{\perp},t_{\parallel},d$, and $f$ are the four strain-susceptibility parameters
that completely describe the strain-response of the orbital states $\vert e_x\rangle$ and $\vert e_y\rangle$.
The effects of these strain components on the electronic states are described by
\begin{eqnarray}
  V_{A_{1g}} &=& \vert e_x\rangle\langle e_x\vert +\vert e_y\rangle\langle e_y\vert \\
  V_{E_{gx}} &=& \vert e_x\rangle\langle e_x\vert -\vert e_y\rangle\langle e_y\vert \\
  V_{E_{gy}} &=& \vert e_x\rangle\langle e_y\vert +\vert e_y\rangle\langle e_x\vert
\end{eqnarray}
The strain Hamiltonian can be rewritten in the basis spanned by the
eigenstates of the spin-orbit coupling
\begin{eqnarray}
H_\text{strain}&=&\hbar\epsilon_{E_{gx}}(L_-+L_+)-i\hbar \epsilon_{E_{gy}}(L_--L_+)
\end{eqnarray}
where $L_+=L_-^\dag=\vert f\rangle\langle g\vert+\vert e\rangle\langle d\vert$ is the orbital raising operator within the ground state.
Upon quantization, the mechanical displacement $\vec{Q}(\vec{r})$
becomes an operator, which  can be expressed
in terms of the elementary normal modes and annihilation operators as
\begin{eqnarray}
\hat{Q}(\vec{r})=\sum_{n,k}[Q_0e^{ikx}\vec{u}_{n,k}(\vec{r})\hat{a}_{n,k}+\text{H.c.}].
\end{eqnarray}
The resulting strain coupling can be written as
\begin{eqnarray}
\hat{H}_\text{strain}\simeq \sum_{n,k}[\hbar g_{n,k}\hat{a}_{n,k}\hat{J}_+e^{ikx_0}+\text{H.c.}],
\end{eqnarray}
with the spin-phonon coupling strength given by
\begin{eqnarray}
  g_{n,k} &=& \frac{d}{v_l}\sqrt{\frac{\hbar\omega_\text{BE}}{2\pi\rho aA}}\varsigma_{n,k}(\vec{r}).
\end{eqnarray}
Here the dimensionless profile is given by
\begin{eqnarray}
\varsigma_{n,k}(\vec{r}) &=& \frac{1}{k}[   (ik u_{n,k}^x+\partial_x u_{n,k}^x- \partial_y u_{n,k}^y+\frac{f}{2d}iku_{n,k}^z+\frac{f}{2d}\partial_x u_{n,k}^z+\frac{f}{2d}\partial_y u_{n,k}^x)\nonumber \\
&&-i (-iku_{n,k}^y-\partial_x u_{n,k}^y-\partial_y u_{n,k}^x+\frac{f}{2d}\partial_y u_{n,k}^z+\frac{f}{2d}\partial_z u_{n,k}^y)].
\end{eqnarray}
For compression modes along the phononic crystal direction $x$, the mode function can
be approximated as $\vec{u}_{n,k}(\vec{r})\sim \vec{e}_x\cos(\omega_{n,k} x/v_l)$, leading to
$\vert \varsigma_{n,k}(\vec{r})\vert =1$.

\subsection{Band-gap interactions between single SiV centers and acoustic modes }

As discussed in the main text, we consider the case where the transition frequency of the
defect center is tuned close to the band edge. Then the spin is dominantly coupled to the modes
near the band edge wavevector $k_0=\pi/a$ due to the van Hove singularity in the density of states, i.e., $\frac{\partial k}{\partial \omega_k}|_{k_0}\rightarrow \infty$.
In this case, the dispersion relation can be approximated to be quadratic $\omega_k\simeq \omega_\text{BE}-\alpha a^2(k-k_0)^2$, with
$\alpha$ a parameter characterizing the band curvature.  In the presence of a static magnetic field
$\vec{B}=B_0\vec{e}_z$ and a microwave driving field of frequency $\omega_0$ and Rabi-frequency $\Omega$ along $\vec{e}_x$, we can
implement a Raman transition between the states $\vert g\rangle$ and $\vert e\rangle$ via the excited state $\vert f\rangle$ through coupling the SiV center to the mechanical modes. In the interaction picture, the full Hamiltonian is given by
\begin{eqnarray}
  \hat{\mathcal {H}} &=& \sum_{k} \hbar g_k \hat{a}_k e^{ikx_0}(\hat{\sigma}_{fg}+\hat{\sigma}_{de})e^{i\delta_\text{BE}t}+\hbar\Omega \hat{\sigma}_{fe}e^{i\delta t}+\text{H.c.}
\end{eqnarray}
In the dispersive limit, i.e., $\delta_\text{BE},\delta\gg g_k,\Omega$, and after adiabatic elimination of the excited states $\vert f\rangle$ and $\vert d\rangle$, we obtain  the effective interaction Hamiltonian
\begin{eqnarray}
\hat{\mathcal {H}} &=&\sum_{k} \hbar g_\text{eff}( \hat{a}_k \hat{\sigma}_{eg}e^{ikx_0}e^{i\Delta_\text{BE}t}+\text{H.c.}),
\end{eqnarray}
with $\Delta_\text{BE}=\Delta-\delta-\omega_\text{BE}$. In the Schr\"{o}dinger picture, we have
\begin{eqnarray}
  \hat{\mathcal {H}} &=& \sum_{k} \hbar \omega_k \hat{a}^\dag_k\hat{a}_k+\hbar \omega_s \hat{\sigma}_{ee}
  +\sum_{k} \hbar g_\text{eff}( \hat{a}^\dag_k
  \hat{\sigma}_{ge}e^{-ikx_0}+\hat{a}_k \hat{\sigma}_{eg}e^{ikx_0}).
\end{eqnarray}

For a single excitation in the system,
there exists a bound state
\begin{eqnarray}
\vert \varphi_b\rangle=\cos \theta\vert 0\rangle\vert e\rangle+\sin\theta\vert 1\rangle\vert g\rangle
\end{eqnarray}
within the bandgap with the eigenenergy $\hbar\Omega_b$. Here $\vert 0\rangle$ is the vacuum state for the phonon modes, and
$\vert 1\rangle=\int dk c_k \hat{a}_k^\dag\vert0\rangle$ is a single phonon excitation of the modes in the band. The bound state and
the corresponding eigenenergy are determined by the eigenvalue equation
\begin{eqnarray}
  \hat{\mathcal {H}}\vert \varphi_b\rangle=\hbar\Omega_b\vert \varphi_b\rangle.
\end{eqnarray}
From this equation we have
\begin{eqnarray}
  \cos\theta(\Omega_b-\omega_s) &=& \sin\theta\sum_kg_\text{eff}c_k e^{ikx_0}\\
  c_k\sin\theta(\Omega_b-\omega_k) &=& g_\text{eff}\cos\theta e^{-ikx_0}.
\end{eqnarray}
Solving these equations  yields
\begin{eqnarray}
  c_k &=& \frac{g_\text{eff}e^{-ikx_0}}{\tan\theta(\Omega_b-\omega_k)} \\
  \Omega_b-\omega_s &=& \sum_k\frac{g_\text{eff}^2}{\Omega_b-\omega_k}\\
  \tan^2\theta&=&\sum_k\frac{g_\text{eff}^2}{(\Omega_b-\omega_k)^2}.
\end{eqnarray}
When we take $\omega_k\simeq \omega_\text{BE}-\alpha a^2(k-k_0)^2$, and change from a discrete modes
to a continuous distribution, we can obtain
\begin{eqnarray}
 \Omega_b-\omega_s &=&\frac{\pi g_\text{eff}^2}{\sqrt{\alpha (\Omega_b-\omega_\text{BE})}} \\
 \tan^2(\theta)&=&\frac{\omega_s-\Omega_b}{2(\omega_\text{BE}-\Omega_b)}\\
 c_k&=& \frac{g_\text{eff}e^{-ikx_0}}{\tan\theta(\Omega_b-\omega_k)}.
\end{eqnarray}
In real space, the phononic part of the bound state is exponentially localized around the spin
with spatial mode envelope
\begin{eqnarray}
\mathcal {E}(x)=\int dk c_k Q_k(x)=\sqrt{\frac{2\pi}{L_{c}}}e^{-\vert x-x_0\vert/L_{c}}Q_{k_0}.
\end{eqnarray}

\subsection{Band gap engineered spin-spin interactions between distant SiV centers}
As discussed in the main text, the band-gap interaction allows us to use the phononic crystal
modes as a quantum bus to perform more complex tasks. Analogous
to the spin-exchange interaction in cavity QED, two  defect centers that are separated by a distance
on the order of the length $L_c$ can exchange a phonon excitation via the induced acoustic cavity mode.
We consider two separated SiV centers coupled to the same phononic crystal modes near the band edge, through
Raman transitions between the states $\vert g\rangle$ and $\vert e\rangle$. In the interaction picture, we can obtain
the interaction between the defect centers and phononic crystal modes
\begin{eqnarray}
 \hat{H}&=& \sum_{j=1,2}\sum_{k}\hbar g_\text{eff}^j( \hat{a}^\dag_k
  \hat{\sigma}_{ge}^je^{-i\delta_k t-ikx_j}+\hat{a}_k \hat{\sigma}_{eg}^je^{i\delta_k t+ikx_j})
\end{eqnarray}
with $\delta_k=\omega_s-\omega_k$. In the dispersive regime $\delta_k\gg g_\text{eff}^j$, this will lead to an effective  spin-spin interaction via the exchange of virtual phonons. With the phononic modes eliminated, the interaction is described by the effective Hamiltonian
\begin{eqnarray}
 \hat{H}&=& \hbar\hat{\sigma}_{eg}^1\hat{\sigma}_{ge}^2g_\text{eff}^2 \int dk \frac{ e^{ik(x_1-x_2)}}{\omega_s-\omega_k}  +\text{H.c.}
\end{eqnarray}
We can integrate over the phononic modes and apply the same approximation near the band edge as used to derive the bound state.
To first order in $\Delta_\text{BE}^{-1}$, the interaction becomes
\begin{eqnarray}
 \hat{\mathcal {H}}_{s-s}&=&\frac{ \hbar g_c^2}{2\Delta_\text{BE}}e^{-\vert x_1-x_2\vert/L_c}\hat{\sigma}_{eg}^1\hat{\sigma}_{ge}^2+\text{H.c.}
\end{eqnarray}

We can generalize the above results to the case where an array of distant defect centers are coupled with the band-gap mediated interaction. This opens another exciting perspective for quantum simulation, with spin-spin Hamiltonians for quantum magnetism of the general form
\begin{eqnarray}
\hat{H}_\text{spin}&=&\sum_{\alpha=x,y,z}\sum_{i,j}J_{i,j}^{\alpha}\hat{\sigma}^{i}_{\alpha}\hat{\sigma}^{j}_{\alpha},
\end{eqnarray}
where $\hat{\sigma}^{i}_{\alpha}$
are Pauli operators and $J_{i,j}^{\alpha}$ are the spin-spin interaction
energies in the $\alpha$ direction for sites $i$ and $j$.
Here, the range of interactions has an exponentially decaying envelope, i.e., $J_{i,j}^{\alpha}\sim e^{-\vert x_i-x_{j}\vert/L_c}$,  and  is restricted
to nearest neighbours.

\end{document}